\documentclass[12pt]{iopart}

\usepackage{graphicx}
\usepackage[usenames]{color}

\begin{document}

\title[LDA+DMFT with pseudopotentials]{LDA+DMFT implemented with the 
pseudopotential plane-wave approach}

\author{G. Trimarchi$^{1,2}$\footnote{Present address: National Renewable Energy 
Laboratory, Golden, CO 80401, USA}, I. Leonov$^1$, N. Binggeli$^{1,2}$, 
Dm. Korotin$^3$, and V. I. Anisimov$^3$}

\address{$^1$   Abdus Salam International Centre for Theoretical Physics,
                         Trieste 34014,
                         Italy}
\address{$^2$   Democritos National Simulation Center, INFM-CNR,
                       Trieste I-34014, Italy }
\address{$^3$   Institute of Metal Physics, Russian Academy of Sciences-Ural Division,
                       620219 Yekaterinburg GSP-170, Russia}


\begin{abstract}

We present a joint implementation of dynamical-mean-field theory (DMFT) with 
the pseudopotential plane-wave approach, via Wannier functions, 
for the determination of the electronic properties of strongly correlated materials. 
The scheme uses, as input for the DMFT calculations, a tight-binding Hamiltonian 
obtained from the plane-wave calculations by projecting onto atomic-centered  
symmetry-constrained Wannier functions for the correlated orbitals.  We apply this 
scheme to two prototype systems: a paramagnetic correlated metal, SrVO$_3$, and a 
paramagnetic correlated system, V$_2$O$_3$, which exhibits a metal-insulator 
transition. Comparison with available 
Linear-Muffin-Tin-Orbital (LMTO) plus DMFT 
calculations demonstrate the suitability of the joint DMFT pseudopotential-plane-wave 
approach to describe the electronic properties of strongly correlated materials. 
This opens the way to future developments using the pseudopotential-plane-wave 
DMFT approach to address also total-energy properties, such as structural properties.

\end{abstract}


\maketitle

\section{Introduction}
Dynamical electron correlations play an important role in the physics of
strongly correlated electronic materials, in particular in determining the  
properties of their paramagnetic and ferromagnetic phases,  as well as their 
metal-insulator transitions. The latter transitions are related to some of the 
most dramatic effects observed in transition-metal oxides, such as the colossal 
magnetoresistance effect in doped manganites \cite{Dagotto05,Tokura00}.
The properties of transition-metal oxides are  also known to be controlled by
a strong and complex interplay between electronic, magnetic, and structural
degrees of freedom \cite{Tokura00}. This interplay leads to giant 
responses to small changes in external parameters such as temperature, pressure, 
magnetic field or doping,  which makes such materials attractive for 
technological applications. 

A new theoretical framework has made possible in recent years the incorporation 
of dynamical correlations in electronic structure calculations of strongly
correlated materials \cite{Kotliar04,Anisimov97,Lichtenstein98}. It combines the 
dynamical-mean-field theory (DMFT) of many-body physics with density-functional 
electronic structure calculations in the local-density approximation (LDA), or in 
the generalized gradient approximation (GGA), and is commonly referred to as 
LDA+DMFT. So far LDA+DMFT computations have been generally implemented   
with the Linear Muffin-Tin Orbital (LMTO) method, and in some cases  also  
with mixed-basis methods \cite{Lechermann06}, which incorporate also 
atomic-like orbitals in the basis sets. In view of the importance, 
however, of the interplay between structural and electronic properties in correlated 
oxides, it would be highly desirable to have a joint implementation of 
DMFT with the electronic-structure pseudopotential plane-wave method. The latter  method 
is indeed known to be well suited to address also total-energy properties,  such as structural 
properties. This is related to the plane-wave basis, in particular, which does not depend on  
the atomic positions in the unit cell and whose completeness can be controlled by a 
single parameter, the plane-wave kinetic energy cutoff. 

Recently, approaches based on Wannier functions  have been proposed to carry 
out LDA+DMFT computations, using e.g., atomic-centered Wannier functions \cite{Anisimov05}, 
or Maximally localized Wannier functions \cite{Lechermann06}, 
or Wannier functions constructed with the N-th order muffin-tin-orbital 
method \cite{Pavarini04}. For strongly correlated materials, 
Wannier functions represent a convenient, physically sound set of localized 
orbitals  for the correlated electrons that can be used to construct an interface 
between  DMFT and LDA/GGA calculations.  
This is especially true for LDA/GGA methods which do not employ 
atomic-like  basis functions, as is the case of the pseudopotential plane-wave method.  

Here we present a joint implementation of DMFT with the pseudopotential 
plane-wave approach, and demonstrate its suitability to determine the electronic properties of 
correlated oxides.  This is a prerequisite to address, in the future,   also 
total-energy and structural properties.  We use as input for the DMFT calculations a 
tight-binding Hamiltonian, H$^{TB} ({\bf k})$,  constructed from the pseudopotential plane-wave 
calculations using atomic-centered symmetry-constrained Wannier functions for the correlated 
orbitals.  We apply this approach to two different test cases: a 
correlated paramagnetic metal, SrVO$_3$, with a simple cubic perovskite structure, 
and a correlated paramagnetic system, V$_2$O$_3$, which has a more complex trigonal crystal 
structure, and exhibits a Mott-Hubbard transition. To our knowledge, this 
is the first successful implementation of LDA+DMFT using the pseudopotential plane-wave 
approach.  

\section{Method}

\subsection{LDA+DMFT calculation scheme}

The standard (although simplified\cite{NoteSCF}) scheme generally used  
to carry out LDA+DMFT calculations includes the following three steps. 

First, an LDA/GGA self-consistent electronic-structure calculation is performed 
for the crystal and phase of interest. The calculated bands associated with the   
relevant correlated orbitals (e.g., the transition-metal $3d$-$t_{2g}$ orbitals 
in our examples) are then mapped onto a 
${\bf k}$-dependent tight-binding Hamiltonian, $\widehat{H}_{LDA}^{TB} ({\bf k})$, 
where ${\bf k}$ is a vector of the crystal Brillouin zone (BZ). 
In practice, in the case of LDA/GGA calculations performed using, as a basis set,  
atomic-like orbitals, this  Hamiltonian is readily obtained as:
\begin{equation} \label{Eq:H_TB}
\widehat{H}_{LDA}^{TB} ({\bf k}) = \sum_{m,m',\sigma} H_{m m'}^{TB (\sigma)}  ({\bf k}) 
 c^{\dagger}_{{\bf k} m \sigma} c_{{\bf k} m' \sigma} 
\end{equation}
where $H_{m m'}^{TB (\sigma)} ({\bf k})$ are the matrix elements, in a given 
${\bf k}$ and $\sigma$ spin-polarization subspace, of the LDA/GGA Hamiltonian 
between the relevant correlated atomic-like orbitals, indexed by $m$, $m'$, and  
$c^{\dagger}_{{\bf k} m \sigma} $ ($c_{{\bf k} m \sigma}$) is the 
creation (annihilation) operator for an electron in orbital $m$ with spin $\sigma$ 
and wavevector ${\bf k}$. 
As we are interested here in the case of paramagnetic phases, we will omit in the 
following   the spin label $\sigma$ in $H_{m m'}^{TB (\sigma)}$. 
We note that, at the LDA/GGA level, the paramagnetic phase is simply modeled by 
a non-magnetic state. 
We are including in the sum, in Eq. (\ref{Eq:H_TB}), a minimal basis 
set of $m$, $m'$ orbitals corresponding to the relevant onsite correlated orbitals 
near the Fermi energy. More generally, however, additional non-correlated orbitals 
may be included,  when needed, to improve the description of the spectrum, e.g., 
further away from the Fermi energy \cite{Kunes07}. 
For simplicity, we will focus in the following on the case of periodic systems 
with a single type of correlated atomic sites.   

Next, the LDA tight-binding Hamiltonian is supplemented with onsite Coulomb 
interactions for the correlated orbitals, in a many-body Hamiltonian of the form: 
\begin{eqnarray}\label{Eq:H_MB}
\widehat{H}& = &\sum_{{\bf k},m,m',\sigma}  
H_{m m'}^{TB} ({\bf k}) c^{\dagger}_{{\bf k} m \sigma} c_{{\bf k} m' \sigma} 
- \sum_{I,m,\sigma} \Delta \epsilon^{dc} n_{I m \sigma}  \nonumber  \\
&+& \sum_{I,m} U n_{I m \uparrow} n_{I m \downarrow}  + 
 \sum_{I,m \neq m',\sigma,\sigma'} (V - \delta_{\sigma \sigma'}J) n_{I m \sigma} n_{I m' \sigma'}. 
\end{eqnarray} 
The index $I$ labels the lattice sites, $m$, $m'$  label the relevant correlated 
atomic orbitals, and $\sigma$, $\sigma'$ the spin states ($\uparrow$ and $\downarrow$), and   
$n_{I m \sigma} $ is the operator for the occupation of the correlated orbitals.   
The first term, on the right-hand side of Eq.(\ref{Eq:H_MB}), is the LDA part of the 
Hamiltonian. The second term is a double-counting correction, formally introduced to 
remove the onsite Coulomb interactions 
already present, in an average way, in the LDA Hamiltonian.  In principle, the 
double-counting potential $\Delta \epsilon^{dc}$ may be taken as \cite{Anisimov97}:  
$\Delta \epsilon^{dc} = U(\bar{n} -\frac{1}{2})$, where $\bar{n}$ is the average occupation  
per correlated orbital. In practice, however, if one takes into account only $d$ orbitals 
(or only $f$ orbitals), the double-counting correction acts on the whole $d$ ($f$) band and 
shifts it by  $\Delta \epsilon^{dc}$. This potential amounts thus to a rigid energy  
shift of the quasiparticle spectrum in the DMFT calculations. It can therefore  
be  absorbed  into the chemical potential, which in turn is determined 
by the number of electrons. Hence the actual value of $\Delta \epsilon^{dc}$ has  
no influence on the spectral properties. The third and forth terms, on the right-hand 
side of Eq.(\ref{Eq:H_MB}), are the interaction terms,  
where $U$ is the onsite Coulomb repulsion parameter, $J$ the Hund's rule exchange parameter, 
and $V = U - 2J$ \cite{Fresard97}.

Finally, the model Hamiltonian, in Eq.(\ref{Eq:H_MB}), is solved by means of DMFT. 
DMFT maps the lattice model onto an effective single-impurity problem subject to a 
self-consistent condition on the impurity self-energy, $\hat{\Sigma} (\epsilon)$, or,  
equivalently, on the local Green's function, $\hat{G} (\epsilon)$ \cite{Georges96}. 
This mapping represents an exact solution in the limit of infinite dimension of the 
lattice problem. The local impurity Green's function and self-energy matrices 
are related through: 
\begin{equation}\label{Eq:Green}
G_{mm'} (\epsilon) = \frac{1}{\Omega_{BZ}} \int d {\bf k} ( [ 
(\epsilon - \mu) {\bf 1} - {\bf H}^{TB} ({\bf k} ) - {\bf \Sigma} (\epsilon) 
]^{-1} )_{m m'}, 
\end{equation}
where $m,m'$ label the correlated orbitals, 
$\mu$ is the chemical potential, ${\bf 1}$, ${\bf \Sigma} (\epsilon)$, and 
${\bf H}^{TB} ({\bf k})$   
are the unitary, self-energy, and LDA tight-binding-Hamiltonian matrices, respectively, 
and the integration extends over the BZ with volume $\Omega_{BZ}$.
Several different approaches may be used to solve the effective impurity problem, 
including numerical renormalization group, exact diagonalization, noncrossing 
approximation, and Quantum Monte Carlo (QMC). In the present work, we employ, as 
impurity solver, the auxiliary fields QMC method by Hirsch and Fye \cite{Hirsch86}. 
For a given input impurity self-energy and Green's function matrices, the impurity 
solver yields a new set of Green's function and self-energy matrices \cite{Georges96}. 
The DMFT equations are solved in an iterative self-consistent cycle, until self-consistency 
is reached.  

\subsection{Tight-binding Hamiltonian from pseudopotential plane-wave calculations}

Wannier functions provide  a physically sound  basis set to construct model Hamiltonians  
for correlated electrons. In the case of LDA/GGA plane-wave calculations,  
they also represent a practical route to build $\hat{H}_{LDA}^{TB} ({\bf k})$, given the 
delocalized nature of the basis functions. 

The Wannier functions are obtained here 
using the approach by Ku {\it et al.} \cite{Ku02}---inspired from the work by Marzari and 
Vanderbilt \cite{Marzari97}, to obtain atomic-centered Wannier functions with 
a given symmetry. First, a set of $M$ trial  functions are generated from the (pseudo) 
atomic wave-functions of the correlated orbitals: 
$|\varphi_{m {\bf k}}\rangle = \sum_{{\bf R}} e^{i {\bf k}{\bf R}} 
|\varphi_m^{{\bf R}} \rangle $, where ${\bf R}$ are the lattice translation 
vectors and  $|\varphi_{m}^{{\bf R}} \rangle$ are the atomic (pseudo) wave-functions,  
in the unit cell ${\bf R}$, with a given symmetry.
A  set of $M$  non-orthogonal Wannier functions (WF) in ${\bf k}$ space 
(or tight-binding Wannier functions) is then obtained by projecting these trial functions 
onto a set of Bloch functions, $ | \psi_{j {\bf k}} \rangle $, belonging to a  
chosen (correlated bands) subspace: 
\begin{equation} \label{Eq:Wannier}
|\widetilde{W}_{m {\bf k}} \rangle = 
\sum_{j=N_1}^{N_2} \langle \psi_{j {\bf k}}| \varphi_{m {\bf k}} \rangle | \psi_{j {\bf k}} \rangle. 
\end{equation}
The sum is over a Bloch subspace defined by imposing either some fixed band numbers,  
$N_1 \leq j \leq N_2$, or an energy window, $E_1 \leq  \epsilon_j ({\bf k}) \leq E_2$, for the 
electronic bands $\epsilon_j ({\bf k})$. 
These Wannier functions are then orthogonalized, by diagonalizing their overlap matrix 
$O_{mm'}$, yielding a set of $M$ orthogonalized Wannier functions:  
$|W_{m {\bf k}} \rangle = \sum_{m'} (O^{-1/2})_{mm'} |\widetilde{W}_{m' {\bf k}} \rangle$. 

Using this basis set of Wannier functions, the tight-binding Hamiltonian matrix is given by: 
\begin{equation} 
H^{TB-WF}_{m m'} ({\bf k}) = \sum_{j=N_1}^{N_2} \langle W_{m {\bf k}} | \psi_{j {\bf k}} \rangle 
\langle  \psi_{j {\bf k}} |  W_{m' {\bf k}} \rangle \epsilon_j ({\bf k}), 
\end{equation}
and can be used, in place of $H^{TB}_{m m'}({\bf k})$, in the many-body Hamiltonian, in 
Eq. (\ref{Eq:H_MB}). It should be noted that the conventional, real-space (lattice-site related) 
Wannier functions,  $ |W_m^{\bf R} \rangle$, are simply the Fourier transforms of the 
tight-binding Wannier functions: 
$ |W_m^{\bf R} \rangle  = 
\frac{1}{N_k} \sum_{{\bf k}} e^{-i {\bf k} {\bf R}} | W_{m {\bf k}} \rangle $.  
Only the tight-binding Wannier functions, however, are explicitly needed in the implementation. 

We implemented this scheme to construct   H$^{TB} ({\bf k})$, in  {\bf k} space,  
in the framework of the pseudopotential plane-wave method. This was done 
as an interface between the  Quantum-Espresso pwscf package \cite{pwscf} and the 
QMC-DMFT code. The matrix elements $\langle \psi_{j {\bf k}}| \varphi_{m {\bf k}} \rangle$, 
in Eq. (\ref{Eq:Wannier}), can be conveniently evaluated in reciprocal space, in the 
pseudopotential plane-wave scheme. For the trial wave-functions 
$|\varphi_{m {\bf k}}\rangle$, we generated wave-functions 
belonging to the point-group representations of the correlated atomic site.  
This was done by diagonalizing the occupation matrix \cite{Cococcioni05} ---calculated in an 
initial (arbitrary) orthogonal basis set in the $l$ angular-momentum subspace of 
the correlated atomic orbitals, and using the corresponding eigenstates, which belong each    
to a specific representation. 
The Hamiltonian H$^{TB} ({\bf k})$ was evaluated on a ${\bf k}$-point grid in the 
irreducible part of the BZ. The integration, in Eq. (\ref{Eq:Green}), was performed  
using the analytical tetrahedron method \cite{Lambin84}, and restricted to the irreducible 
part of the BZ by symmetrizing the Green's function matrix. 

\subsection{Technical details: QMC-DMFT computations}

Computationally, the most involving part of the calculations is the evaluation of the 
path integral in the auxiliary-field QMC method, to solve for the local impurity Green's 
function \cite{Georges96}. The QMC method maps the interacting electron problem onto 
a sum of non-interacting problems by means of the Trotter discretization  and 
Hubbard-Stratonovich transformation \cite{Georges96,Held02}. 
The imaginary time integrals are represented, in the 
Trotter discretization, by  $L$ imaginary-time slices of size $\Delta  = \beta/L$,  with   
$\beta =  1/K_B T$. For an $M$-orbital impurity problem, the Hubbard-Stratonovich 
transformation introduces  $M (2M-1)$  auxiliary Ising fields for each time slice. 
In addition to the tolerance parameter of the DMFT self-consistency, the number of 
time slices $L$ and the number of Monte Carlo sweeps $N_{MC}$ for the stochastic integration 
of the path integral are the sole convergence parameters of the QMC-DMFT calculations. 
It should be noted that the computational cost of the QMC algorithm scales, to the leading 
order in $L$, as $\sim M (2M-1) N_{MC} L^3$ \cite{Held02}. 

The QMC method has the advantage 
of having formally no approximation. Very low temperatures, however, are not accessible, 
as the numerical effort scales as $1/T^3$. 
From the imaginary-time self-consistent Green's function  obtained from the 
QMC-DMFT computations, the real-frequency single particle spectral functions are computed   
using the maximum entropy method \cite{Jarrell96}. 

\section{Applications}

The joint DMFT pseudopotential plane-wave scheme described in the previous sections, 
was applied to two test cases, SrVO$_3$ and V$_2$O$_3$. For the density functional 
calculations, we used the  Perdew-Burke-Ernzerhof exchange-correlation functional \cite{pbe} 
together with Vanderbilt utrasoft pseudopotentials \cite{Vanderbilt}.  We used a kinetic energy 
cutoff of  35 Ry (350 Ry)  for the plane-wave expansion of the electronic 
states (core-augmentation charge).  The self-consistent calculations were performed with a 
(4,4,4) Monkhorst-Pack ${\bf k}$-point grid \cite{MP}. For the computation of ${\bf H}^{TB} ({\bf k}) $ 
and of the Green's functions matrix, in Eq. (\ref{Eq:Green}), we used a (10,10,10) ${\bf k}$-point 
grid centered at $\Gamma$.  The experimental values of the lattice parameters of SrVO$_3$ 
($a = 3.84$ \AA) \cite{Reya90} and V$_2$O$_3$ ($a = 4.95$ \AA, $c = 14.00$ \AA) \cite{Keller04} 
have been used in our calculations. 

For SrVO$_3$, we set the onsite Coulomb interaction U = 5.55 eV and Hund's rule parameter 
 J = 1 eV \cite{Nekrasov05}. 
For V$_2$O$_3$, we used $J = 0.93$ eV \cite{Keller04} and several different values of 
$U$ \cite{Anisimov05}. 
The QMC-DMFT calculations were performed at $T = 580$ K 
($\beta = 20$ eV$^{-1}$), using 80 imaginary-time slices. In the case of V$_2$O$_3$, we also 
performed calculations at  $T = 1160$ K ($\beta = 10$ eV$^{-1}$), using 40 imaginary-time 
slices. In all QMC-DMFT calculations we used $\sim 10^6$ Monte Carlo sweeps. 

\subsection{SrVO$_3$}

\begin{figure}[htb]
\vspace*{1.5cm}
\begin{center}
\includegraphics[width=0.9\textwidth]{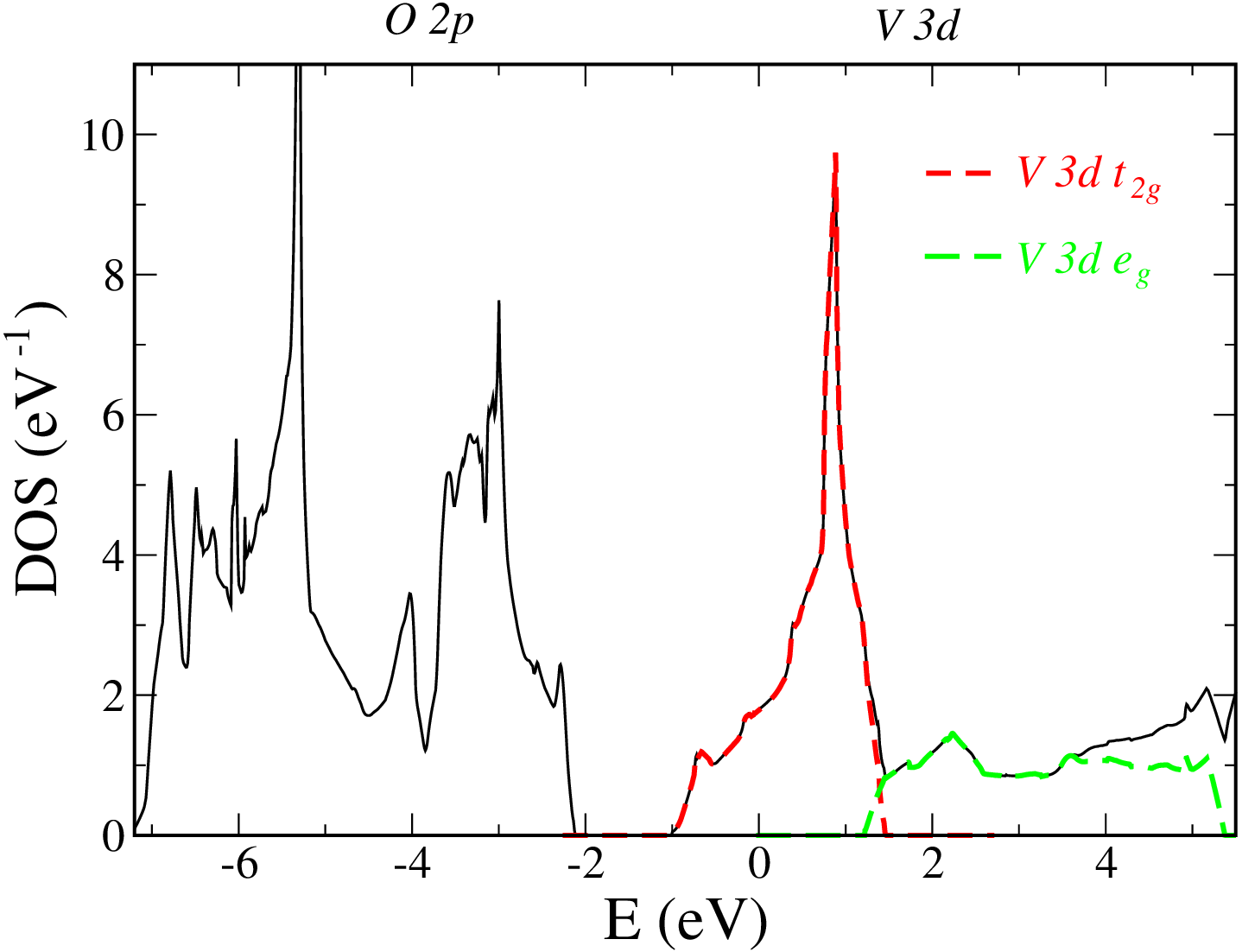} 
\end{center}
\caption{(Color online) Density of states of SrVO$_3$ obtained from the GGA pseudopotential 
plane-wave calculations. The short- (long-) dashed line shows the V-$3d$- $t_{2g}$ 
($e_g$) Wannier-projected density of states. The generic characters of the bands are  
also indicated at the top of the figure. The zero of energy corresponds to the 
Fermi level.}
\label{fig:SrVO3_DOS}
\end{figure}

SrVO$_3$ is a prototype $d^1$ correlated paramagnetic metal. It has a simple cubic 
perovskite structure, and remain paramagnetic down to low temperatures.  
It is   an ideal test case for first-principles many-body calculations.  
In Fig. \ref{fig:SrVO3_DOS}, we show the density of states (DOS) obtained for SrVO$_3$ 
from the GGA pseudopotential calculations. The spectrum is in agreement with previous 
LDA calculations \cite{Lechermann06,Nekrasov05}. The valence states of SrVO$_3$ consist 
of completely occupied oxygen $2p$ states, located in the energy range -7 eV to 
-2 eV below the Fermi energy, and partially occupied V-$3d$ $t_{2g}$ states, 
near the Fermi energy. 

From the pseudopotential plane-wave calculations, we generated the  
V-$3d$-  $t_{2g}$  and $e_{g}$ Wannier functions from the Bloch functions corresponding to 
the 5 lowest-energy ($3d$) bands, in the energy window -1 to 5.5 eV. The corresponding projected 
Wannier $t_{2g}$  and $e_{g}$ DOS's are also displayed in Fig. \ref{fig:SrVO3_DOS}. 
Because of the ideal octahedral 
symmetry of the V sites, hybridization is forbidden between the  $t_{2g}$  and $e_{g}$ 
states.  ${\bf H}^{TB} ({\bf k} )$ is hence block diagonal with respect to these two 
subspaces. 
We have used, as correlated subspace, the $t_{2g}$ Wannier-functions subspace, 
and the corresponding 
$t_{2g}$-Hamiltonian block for the DMFT computations.  

It should be noted that, in the special case of cubic symmetry, the Green's 
function matrix of the $t_{2g}$ states, in Eq. (\ref{Eq:Green}), may be expressed as: 
$G_{mm'} (\epsilon) =  \int \frac{d \epsilon  D(\epsilon')} 
{\epsilon - \mu -\epsilon' - \Sigma_{m m}(\epsilon')} \delta_{mm'}$. This is valid, 
however, only when  the local matrices (self-energy and Green's function matrices)  
are proportional to the unitary matrix, i.e., in the case of cubic and higher symmetry. 
In the present work we always used the more general Hamiltonian formulation with  the 
${\bf k}$ space integration. 

\begin{figure}[htb]
\vspace*{5cm}
\begin{center}
\includegraphics[width=0.6\textwidth]{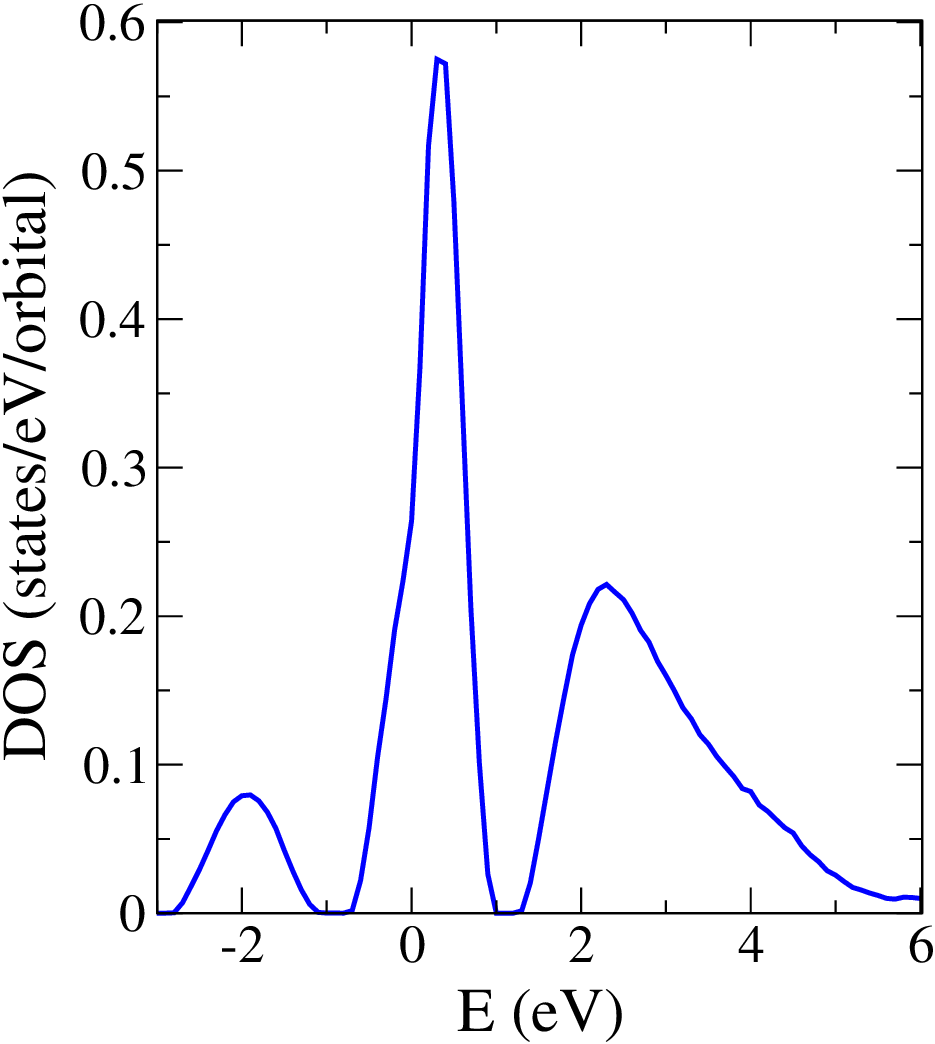} 
\end{center}
\caption{(Color online) Spectral function for SrVO$_3$ at $T = 580$ K obtained from the GGA+DMFT 
computations using the $t_{2g}$ projected Hamiltonian from the pseudopotential 
plane-waves  calculations. The zero of energy corresponds to the Fermi level. }
\label{fig:SrVO3_DMFT}
\end{figure}

In Fig. \ref{fig:SrVO3_DMFT}, we display the corresponding single-particle spectral 
function obtained from the QMC-DMFT calculations at $T = 580$ K. Taking into account 
the correlations effects within the $t_{2g}$ manifold leads to substantial modifications  
in the single-particle spectrum relative to the GGA result. Correlations 
effects are responsible for a lower Hubbard band around -2 eV, an upper Hubbard 
around 2.5 eV, and a well-pronounced quasiparticle peak at the Fermi energy.
This is in general agreement with the photoemission and inverse photoemission 
experiments on SrVO$_3$  \cite{Sekiyama04}.   
The results in Fig. \ref{fig:SrVO3_DMFT} compare  well with previous 
LMTO-based LDA+DMFT computations performed with the same value of U \cite{Nekrasov05}, and  are 
also consistent with the mixed-basis LDA+DMFT calculations using somewhat smaller 
values of U \cite{Lechermann06}. 

\subsection{V$_2$O$_3$} 

V$_2$O$_3$  is a Vanadium $d^2$ system.  
The high-temperature paramagnetic phase of V$_2$O$_3$ has a corundum trigonal crystal 
structure, with  4 equivalent V sites in the unit cell. 
Within the corundum structure, each V ion is surrounded by a distorted oxygen 
octahedron \cite{Tanaka02,Hebert02}.  
The V ions are arranged in pairs along the $c$ axis, with a stacking that  
can be obtained, starting from an ideal chain of V ions  along ${\bf c}$, by 
introducing vacancies at every third site \cite{Keller04}. 
 While the V-V pairs along the $c$ axis are 
surrounded by face-sharing oxygen octahedra, in the ${\bf a}$-${\bf b}$ plane, each 
V ion has three nearest neighbors with edge-sharing oxygen octahedra \cite{Tanaka02}.  

Assuming ideal V octahedral sites in V$_2$O$_3$, the V-$3d$ atomic states are split 
into   $t_{2g}$ and   $e_g$ orbitals, where the two degenerate $e_g$ orbitals are empty
and the three degenerate $t_{2g}$  orbitals are filled with 2 electrons.  When the symmetry 
is further reduced by a small trigonal distortion of the octahedra in the corundum structure, 
the $t_{2g}$ orbitals further splits into a non-degenerate $a_{1g}$ orbital oriented along 
the $c$ axis, and two degenerate  $e_{g}^{\pi}$ orbitals oriented predominantly in the  
${\bf a}$-${\bf b}$ plane.  

\begin{figure}[htb]
\vspace*{2cm}
\begin{center}
\includegraphics[width=0.9\textwidth]{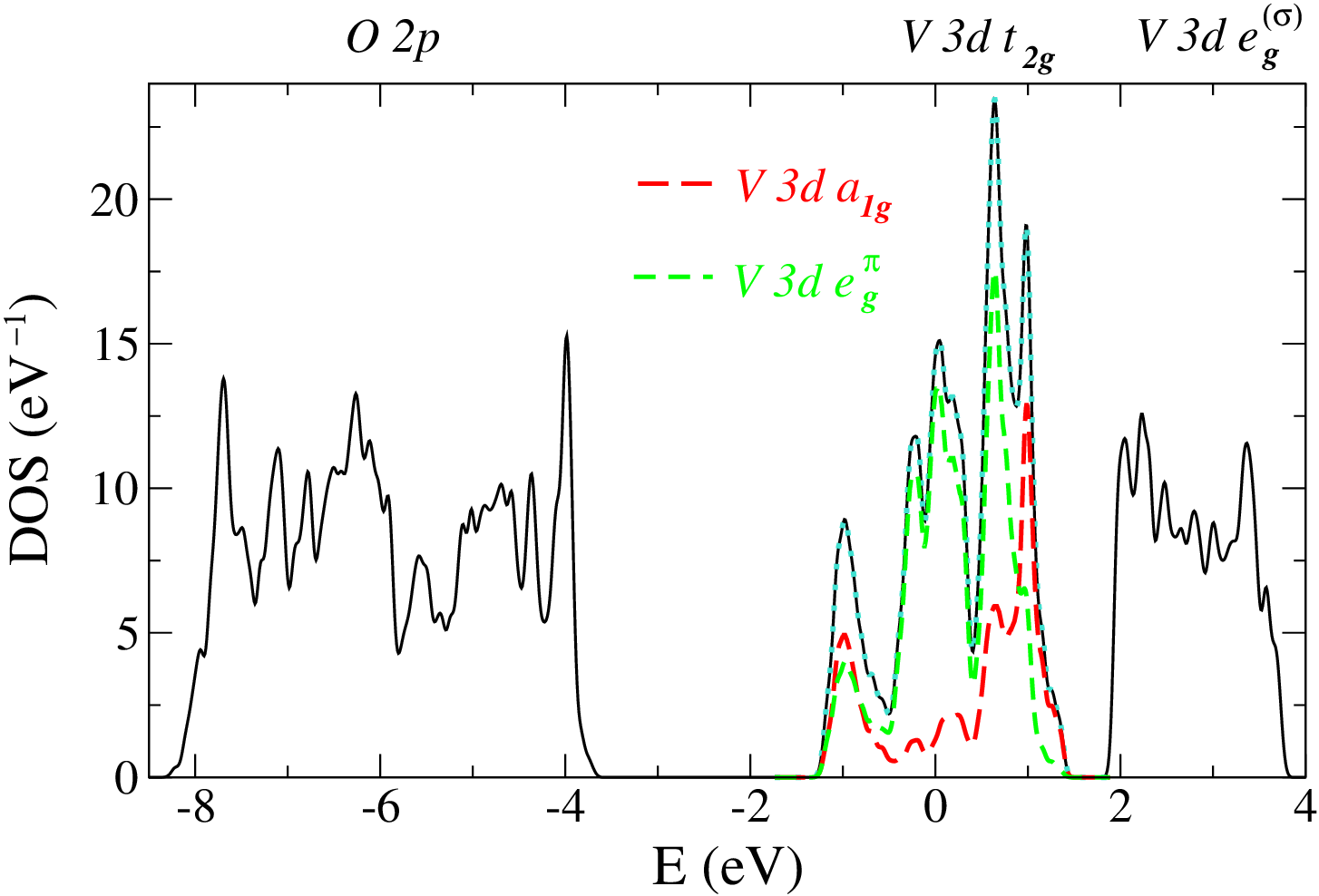} 
\end{center}
\caption{(Color online) Density of states of V$_2$O$_3$ obtained from the GGA pseudopoential 
plane-waves calculations. The short- (long-) dashed line shows the V-$3d$-  
$e_g^{\pi}$ ($a_{1g}$) Wannier-projected density of states. The generic characters 
of the bands are  also indicated at the top of the figure. 
The zero of energy corresponds to the Fermi level. }
\label{fig:V2O3_DOS}
\end{figure}

In Fig. \ref{fig:V2O3_DOS}, we show the DOS of V$_2$O$_3$ obtained 
from the GGA pseudopotential calculations. 
The oxygen $2p$ states of V$_2$O$_3$ are located roughly between -8 and -4 eV below 
the Fermi energy. 
The V-$3d$ $t_{2g}$-like states are located in an energy window between -1.5 eV and 
1.5 eV, and the $e_g^{\sigma}$  states are approximatively between $\sim 2$ eV and $4$ eV.
This is consistent with previous LDA calculations \cite{Keller04,Hebert02,Held01}. 
The GGA (and also LDA) calculations yield a metallic phase for V$_2$O$_3$, 
with a high DOS at the Fermi energy. Experimentally, instead, the paramagnetic  
corundum phase is found to be an insulator at low pressure. 

\begin{figure}[htb]
\vspace*{1.5cm}
\begin{center}
\includegraphics[width=0.9\textwidth]{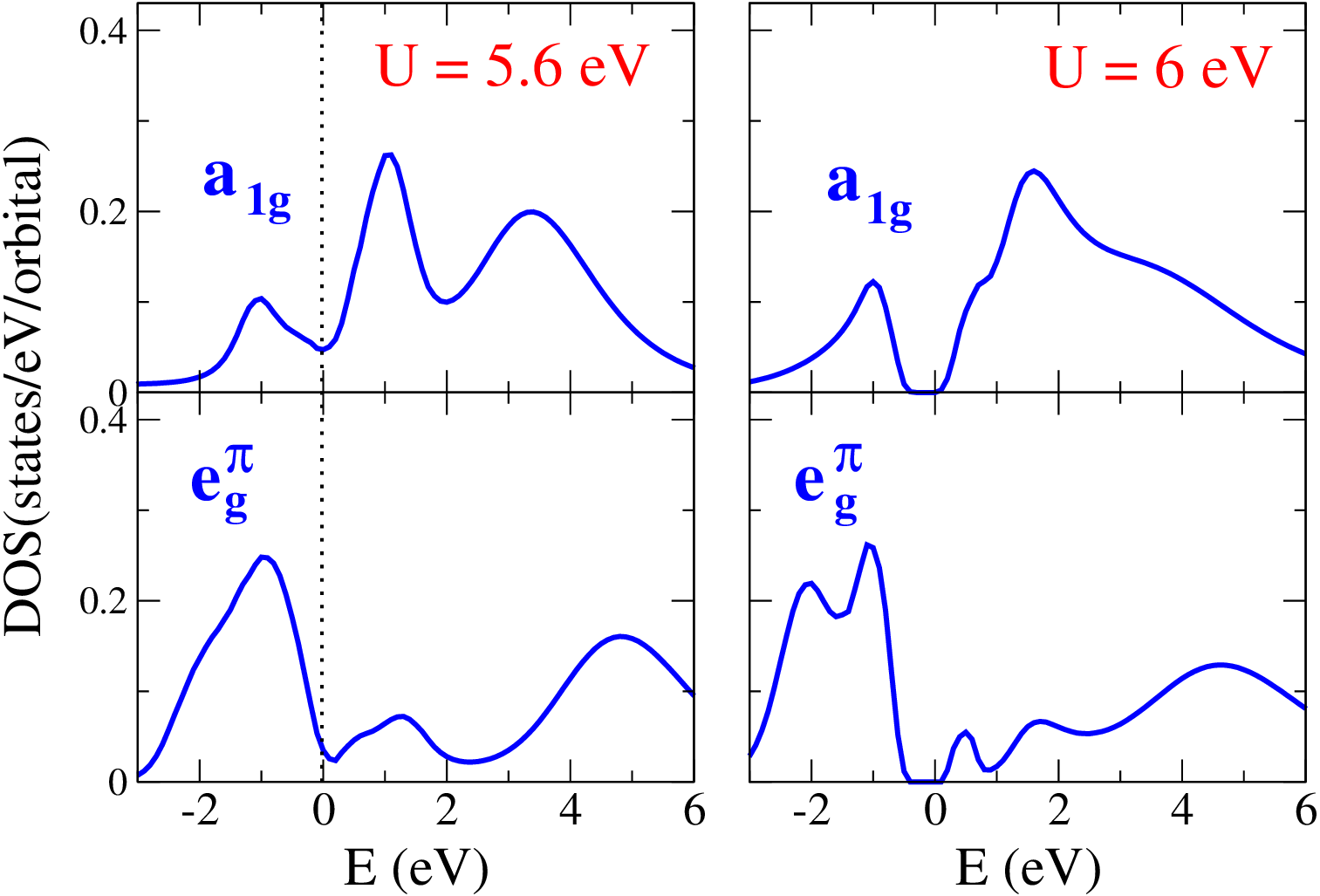} 
\end{center}
\caption{(Color online) V-$3d$ $a_{1g}$ and $e_g^{\pi}$ contributions to the spectral 
function of V$_2$O$_3$ at $T = 1160$ K, as obtained from the GGA+DMFT 
computations using the $t_{2g}$-like Hamiltonian constructed from the 
pseudopotential plane-waves  calculations. The results are shown for two 
different values of the onsite Coulomb interaction, $U = 5.6$ and 
$6$ eV. The vertical dotted line indicates the Fermi energy.}
\label{fig:V2O3_DMFT_1160K}
\end{figure}

\begin{figure}[htb]
\vspace*{1.5cm}
\begin{center}
\includegraphics[width=0.9\textwidth]{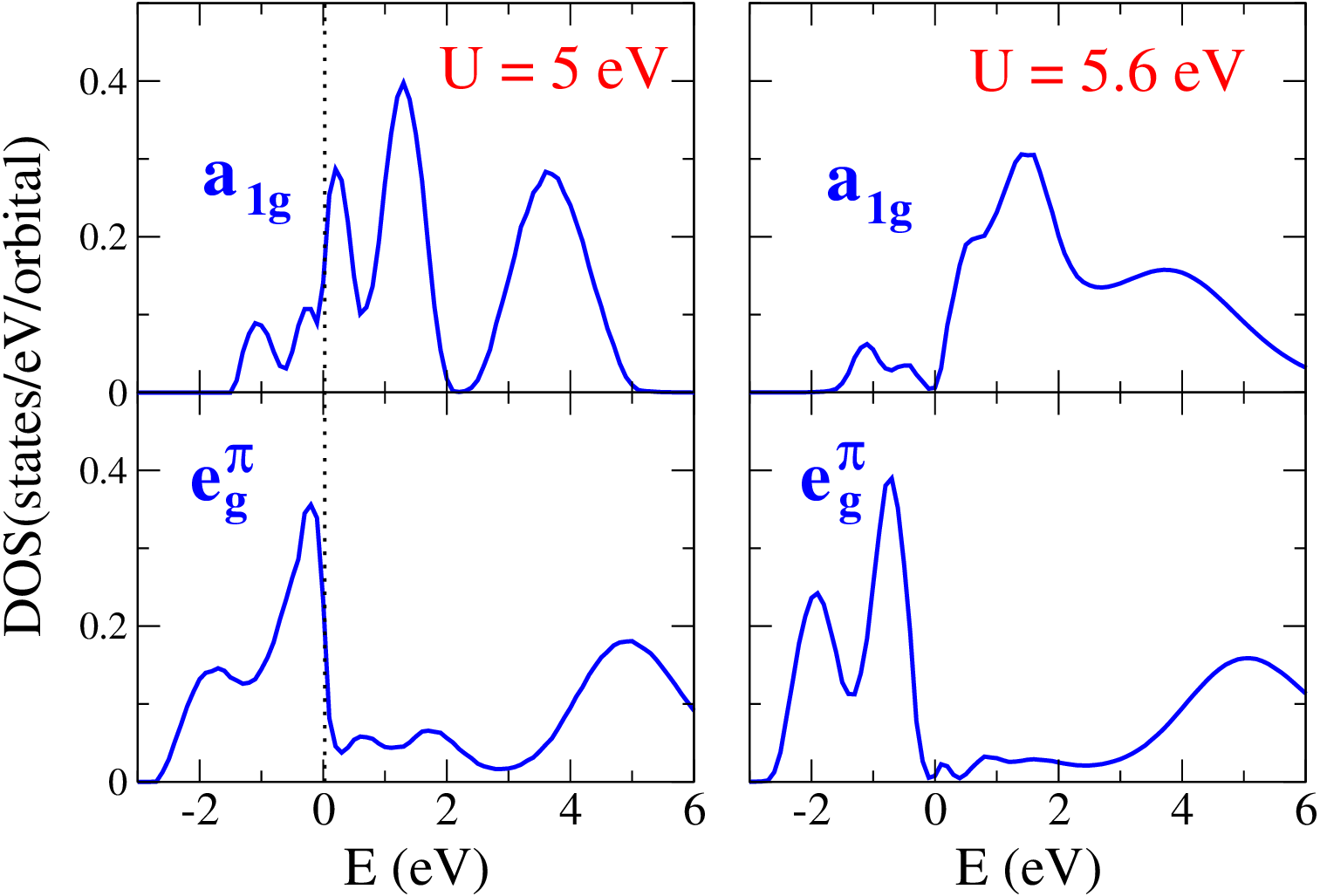} 
\end{center}
\caption{(Color online) Same data as in Fig. 4, but for $T = 580$ K and $U = 5$ and $5.6$ eV.}
\label{fig:V2O3_DMFT_580K}
\end{figure}

We generated Wannier functions 
with $a_{1g}$ and $e_g^{\pi}$ symmetry from the Bloch states enclosed in the energy 
window  -1.5 to 1.5 eV (see Fig.  \ref{fig:V2O3_DOS}). The corresponding Wannier 
projected DOS's are  displayed in Fig.  \ref{fig:V2O3_DOS}. We used these 
Wannier functions to construct ${\bf H}^{TB} ({\bf k})$  and carry out the GGA+DMFT computations. 
The results are shown in Figs. \ref{fig:V2O3_DMFT_1160K} 
and \ref{fig:V2O3_DMFT_580K} for two different temperatures: $T = 1160$ K and 
580 K, respectively. The separate $a_{1g}$ and $e_g^{\pi}$ contributions to the spectral 
function are displayed in the upper and lower panels, respectively. At T = 1160 K,   
calculations were performed for $U = 5.6$ and $6$ eV. An insulating phase is obtained for  
$U = 6$ eV, while the system is still metallic at U = 5.6 eV. The results in Fig. 
 \ref{fig:V2O3_DMFT_1160K}  agree well with the available LMTO-based LDA+DMFT 
calculations performed at the same temperature \cite{Anisimov05,Keller04,Held01}, and 
in particular with the results obtained using a Wannier-projected 
Hamiltonian \cite{Anisimov05}. 
The pseudopotential implementation is found to yield a slightly larger critical U for the 
insulating phase ($U \sim 6$ eV) compared to the LMTO implementation ($U \sim 5.5$ eV) 
\cite{Anisimov05}. This is related to the band width of the LDA/GGA $t_{2g}$-like states, 
which is slightly larger in the pseudopotential case compared to the LMTO case. 

At $T = 580$ K, we performed calculations for $U = 5$ and $5.6$ eV. An insulating phase is 
found at $U \sim 5.6$ eV, whereas at $U = 5$ eV the system is metallic, with a large DOS 
at the Fermi energy. In the latter case, one observes a quasi-particle peak at the Fermi 
energy in the $a_{1g}$ orbital-resolved spectral function. The results at $T = 580$ K,  
in Fig. \ref{fig:V2O3_DMFT_580K}, are in good qualitative agreement with the results of LDA+DMFT calculations using a 
Wannier Hamiltonian constructed with the N-th order muffin-tin-orbital method \cite{Poteryaev07}. 
In the latter study, a critical $U$ of 4.2 eV was found for the insulating phase. The larger  
critical $U$ obtained here  (5.6 eV) is attributed mainly to the 
difference in the crystal structure; in Ref. \cite{Poteryaev07} 
the (V$_{0.962}$Cr$_{0.038}$)$_2$O$_3$ 
crystal structure was considered ---which corresponds to the experimental insulating structure. 
The increased $U$ is also due in part 
to the broader GGA band found in the pseudopotential calculations. 
One observes also some small differences in the peak structure, between the present results 
in Fig. \ref{fig:V2O3_DMFT_580K}  and the spectral function in Ref. \cite{Poteryaev07}. 
These are attributed mainly to differences in the DMFT calculational details, such as the 
use of two different interpolation schemes (Ulmke-Janis-Vollhardt 
scheme \cite{Ulmke95} versus cubic splines) for the Fourier transformation of the local 
Green's function, in the two studies.  

\subsection{Discussion and outlook} 

The results we have presented here show that the  DMFT plus pseudopotential plane-wave 
scheme is both a practical and suitable approach for the determination of the electronic 
properties of strongly correlated oxides. A promising extension of this 
approach concerns the determination of total-energy properties,  
and in particular of structural properties of correlated systems. The total 
energy, within the LDA+DMFT, can be expressed as \cite{Amadon06}: 
$E = E_{LDA} - E_{DC}  - \sum_{m, {\bf k}} \epsilon_{m, {\bf k}}^{LDA} + 
\langle  H^{TB}_{LDA} \rangle + 
\langle H_{U} \rangle $, where $E_{LDA}$ is the LDA 
total energy, $E_{DC}$ is the double-counting energy corresponding to the second 
term on the right-hand side of Eq. (\ref{Eq:H_MB}), 
$\sum_{m, {\bf k}} \epsilon_{m, {\bf k}}^{LDA} $ is the sum of the LDA 
valence-state eigenvalues, $\langle  H^{TB}_{LDA} \rangle 
= {\rm tr} [{\bf H}^{TB} {\bf G}] $, and $\langle H_{U} \rangle$ is the interaction 
energy, corresponding to the third and fourth term on the right-hand side of 
Eq. (\ref{Eq:H_MB}), computed from the double occupancy matrix \cite{McMahan05}. 
The application of the DMFT pseudopotential plane-wave approach 
to the determination of structural relaxations is  a line of 
development we are currently pursuing. 

\section{Summary and conclusions}

We have presented an implementation of the LDA+DMFT approach within the pseudopotential 
plane-wave framework. This scheme was applied to two different test cases, SrVO$_3$ 
and V$_2$O$_3$. Comparison with available LMTO-based LDA+DMFT calculations   
demonstrated  the suitability of the joint DMFT pseudopotential-plane-wave 
scheme to describe the electronic properties of strongly correlated materials.
This opens the way to future developments using the DMFT 
pseudopotential-plane-wave approach to address also total-energy and hence 
structural properties of correlated systems.

\ack
We thank  S. de Gironcoli, M. Altarelli, and D. Vollhardt for helpful discussions.  
We are also grateful to D. Vollhardt for providing the QMC-DMFT code. We acknowledge 
support for this work by the Light Source Theory Network, LighTnet, of 
the EU. Calculations were performed on the IBM sp5 computer at CINECA.


\begin{thebibliography}{10}
\bibitem{NoteSCF} LDA+DMFT calculations are presently performed using a non 
selfconsistent scheme. As has been proposed in Refs. \cite{Anisimov05} and 
\cite{Lechermann06}, however, the charge density could be recalculated from the 
DMFT solution (Green's function) and injected into an iterative LDA+DMFT loop, 
to reach a self-consistent solution. In practice, however, 
this is a very demanding procedure, which to the best of our knowledge has not 
been successfully implemented so far.
\end{thebibliography}
\end{document}